\documentclass[sigconf]{acmart}

\AtBeginDocument{%
  \providecommand\BibTeX{{%
    \normalfont B\kern-0.5em{\scshape i\kern-0.25em b}\kern-0.8em\TeX}}}

\copyrightyear{2024}
\acmYear{2024}
\setcopyright{rightsretained}
\acmConference[MM '24]{Proceedings of the 32nd ACM International Conference on Multimedia}{October 28-November 1, 2024}{Melbourne, VIC, Australia}
\acmBooktitle{Proceedings of the 32nd ACM International Conference on Multimedia (MM '24), October 28-November 1, 2024, Melbourne, VIC, Australia}
\acmDOI{10.1145/3664647.3681620}
\acmISBN{979-8-4007-0686-8/24/10}

\usepackage{pifont}
\usepackage{booktabs}
\usepackage{siunitx}
\begin{document}
\newcommand{\name}{PSM}
\newcommand{\cmark}{\ding{51}}%
\newcommand{\xmark}{\ding{55}}%
\title{{\name}: Learning Probabilistic Embeddings for Multi-scale Zero-Shot Soundscape Mapping}


\author{Subash Khanal}
\affiliation{%
  \institution{Washington University in St. Louis}
  \city{Saint Louis}
  \country{United States}}
\email{k.subash@wustl.edu}

\author{Eric Xing}
\affiliation{%
  \institution{Washington University in St. Louis}
  \city{Saint Louis}
  \country{United States}}
\email{e.xing@wustl.edu}

\author{Srikumar Sastry}
\affiliation{%
  \institution{Washington University in St. Louis}
  \city{Saint Louis}
  \country{United States}}
\email{s.sastry@wustl.edu}

\author{Aayush Dhakal}
\affiliation{%
  \institution{Washington University in St. Louis}
  \city{Saint Louis}
  \country{United States}}
\email{a.dhakal@wustl.edu}

\author{Zhexiao Xiong}
\affiliation{%
  \institution{Washington University in St. Louis}
  \city{Saint Louis}
  \country{United States}}
\email{x.zhexiao@wustl.edu}

\author{Adeel Ahmad}
\affiliation{%
  \institution{Washington University in St. Louis}
  \city{Saint Louis}
  \country{United States}}
\email{aadeel@wustl.edu}

\author{Nathan Jacobs}
\affiliation{%
  \institution{Washington University in St. Louis}
  \city{Saint Louis}
  \country{United States}}
\email{jacobsn@wustl.edu}
\renewcommand{\shortauthors}{Khanal et al.}

\begin{abstract}

A soundscape is defined by the acoustic environment a person perceives at a location. In this work, we propose a framework for mapping soundscapes across the Earth. Since soundscapes involve sound distributions that span varying spatial scales, we represent locations with multi-scale satellite imagery and learn a joint representation among this imagery, audio, and text. To capture the inherent uncertainty in the soundscape of a location, we design the representation space to be probabilistic. We also fuse ubiquitous metadata (including geolocation, time, and data source) to enable learning of spatially and temporally dynamic representations of soundscapes. We demonstrate the utility of our framework by creating large-scale soundscape maps integrating both audio and text with temporal control. To facilitate future research on this task, we also introduce a large-scale dataset, GeoSound, containing over $300k$ geotagged audio samples paired with both low- and high-resolution satellite imagery. We demonstrate that our method outperforms the existing state-of-the-art on both GeoSound and the existing SoundingEarth dataset. Our dataset and code is available at \url{https://github.com/mvrl/PSM}.
\end{abstract}





\maketitle

\section{Introduction}
\label{sec:intro}


Soundscape mapping involves understanding the relationship between locations on Earth and the distribution of sounds at those locations. The soundscape of an area strongly correlates with psychological and physiological health ~\cite{lercher2023soundscape}. Therefore, soundscape maps are valuable tools for environmental noise management and urban planning ~\cite{margaritis2017soundscape,gonzalez2023effects,ooi2023araus}. Additionally, technologies such as augmented reality and navigation systems can use soundscape mapping to provide immersive experiences.


Traditionally, soundscape mapping involves predicting a set of acoustic indicators (e.g., sound pressure, loudness) to descriptors (e.g., pleasant, eventful) ~\cite{international2014iso,lionello2020systematic,engel2021review}. This approach limits understanding of the acoustic scene and relies on crowd-sourced data ~\cite{picaut2019open,aiello2016chatty}, often available only for densely populated areas. As a result, traditional methods generate sparse soundscape maps that lack generalizability and are unsuitable for creating global maps.

To address the limitations of traditional soundscape mapping, we adopt a formulation where, given a specific location, the task is to train a machine learning model that directly predicts the sound distribution likely to be encountered at that location. We represent each location with a satellite image centered around it. This approach enables the generalization of soundscape mapping beyond locations explicitly included in the training data.

We approach the soundscape mapping problem from the perspective of multimodal representation learning to design a shared embedding space between audio and satellite imagery at the recorded location of the audio. This learning strategy aims to bring positive audio-satellite image pairs closer while pushing negative pairs farther apart in the embedding space. Ultimately, the multimodal embedding space can be employed to generate soundscape maps by computing similarity scores between the query and the satellite image set covering the geographic region of interest.


Soundscape mapping is inherently uncertain, as multiple types of sounds can come from one location, and a specific sound can originate from multiple locations. Paired location and audio data often contain pseudo-positives, sample pairs labeled as negatives but semantically similar to positives. Methods that learn deterministic representations of sound and satellite imagery ignore this uncertainty. To address this, we propose a probabilistic multi-modal embedding space for audio, satellite imagery, and textual descriptions ~\cite{chun2023improved}. To account for potential false negatives, we identify pseudo-positive matches during training ~\cite{chun2023improved}.


Satellite images of audio capture locations can be obtained at different spatial resolutions, with ground area coverage increasing as zoom levels increase. To create large-scale soundscape maps, we aim to learn an embedding space that models these differences in spatial resolution of satellite imagery. Therefore, we modify zero-shot soundscape mapping to multi-scale zero-shot soundscape mapping, allowing ground-level sounds to be mapped with satellite imagery at different zoom levels. We achieve this by learning a shared satellite image encoder across zoom levels using Ground-Sample Distance Positional Embedding (GSDPE)~\cite{reed2023scale}.

Our modalities of interest, satellite imagery, audio, and text, often have associated metadata that convey meaningful information (such as latitude and longitude or the source of an audio sample). We propose to fuse such metadata: location, time, and source from which the audio was collected, into our framework. We demonstrate that such information increases the discriminative power of our embedding space and allows the creation of soundscape maps conditioned on dynamic metadata settings during inference. 


The most closely related prior work \cite{khanal2023soundscape} in soundscape mapping was trained on limited data ($\sim35$k samples) from the \textit{SoundingEarth} dataset \cite{heidler2023self}. To advance research in this area, we curated a new large-scale dataset, \textit{GeoSound},collecting geotagged audios from four different sources, we increased the dataset size by six-fold to over $300$k samples. We use \textit{GeoSound} to train our framework that advances the state-of-the-art in zero-shot soundscape mapping by learning a probabilistic, scale-aware, and metadata-aware joint multimodal embedding space. Moreover, we demonstrate the capability of the proposed framework in the creation of temporally dynamic soundscape maps.


The main contributions of our work are as follows:
\begin{itemize}
  \item We introduce a new large-scale dataset containing over $300$k geotagged audios paired with high-resolution ($0.6$m) and low-resolution ($10$m) satellite imagery.
  \item We learn a metadata-aware, probabilistic embedding space between satellite imagery, audio, and textual audio description for zero-shot multi-scale soundscape mapping.
  \item We demonstrate the utility of our framework ({\name}: \underline{\textbf{P}}robabilistic \underline{\textbf{S}}oundscape \underline{\textbf{M}}apping) in creating large-scale soundscape maps created by querying our learned embedding space with audio or text.
\end{itemize}

\section{Related Works}
\label{sec:litreview}

\subsection{Audio Visual Learning}
An intricate relationship exists between the audio and visual attributes of a scene. Utilizing this relationship, there have been numerous works in the field of audio-visual learning.~\cite{owens2016ambient,salem2018multimodal,zhao2023sensing,hu2020cross,heidler2023self,khanal2023soundscape, SpecVQGAN_Iashin_2021,zeng2023learning,cheng2020look}. Owens \textit{et al.}~\cite{owens2016ambient} have demonstrated that encouraging the models to predict sound characteristics of a scene allows them to learn richer representations useful for visual recognition tasks. Hu \textit{et al.}~\cite{hu2020cross} proposed to learn from audio and images to solve the task of aerial scene recognition. Relatively closer to the formulation of our work, Salem \textit{et al.}~\cite{salem2018multimodal} proposed to learn a shared feature space between satellite imagery, ground-level concepts, and audio, which allowed them to predict sound cluster distribution across large geographic regions. Recently, Khanal \textit{et al.}~\cite{khanal2023soundscape} proposed the learning of a tri-modal embedding space to map satellite imagery with the most likely audio at a location. 


\subsection{Deterministic Contrastive Learning}
The contrastive learning paradigm \cite{radford2021learning,10.1145/3503161.3548263,ma2022x,tao2020self} has significantly advanced state-of-the-art multimodal learning capabilities through rich cross-modal supervision. In the pursuit of advancing contrastive learning approaches for audio and text, Elizalde et al. \cite{elizalde2022clap} and Wu et al. \cite{laionclap2023} have developed a Contrastive Language-Audio Pretraining (CLAP) framework, showcasing strong zero-shot capabilities. Wav2CLIP \cite{wu2022wav2clip} distills information learned from CLIP to create a joint image-audio embedding space. AudioCLIP \cite{guzhov2022audioclip} extends contrastive learning to audio, image, and text, exhibiting impressive performance across various downstream tasks. Recently, Heidler et al. proposed learning a shared representation space between audio and corresponding satellite imagery for use in various downstream tasks in remote sensing. Similarly, Khanal et al. \cite{khanal2023soundscape} utilized the \textit{SoundingEarth} dataset \cite{heidler2023self} to train a multimodal embedding space using a deterministic contrastive loss \cite{infonce} and used this model for zero-shot soundscape mapping.

\subsection{Probabilistic Contrastive Learning}
In our formulation of soundscape mapping, the satellite image provided as location context captures a geographic area containing many sound sources. As such, deterministic contrastive learning approaches cannot capture the inherent ambiguity in the mapping from satellite image to sound, as any sample can only be represented by a single point in the embedding space. This limitation can be addressed by representing embeddings probabilistically ~\cite{aljundi2023contrastive, ji2023map, 9554405,huang2021learning,upadhyay2023probvlm,neculai2022probabilistic,chun2021probabilistic,chun2023improved,su2021modeling}. In other words, each sample in probabilistic embedding space is represented by a probability distribution whose parameters are learned, usually by a neural network.


Probabilistic Cross-Modal Embeddings (PCME) ~\cite{chun2021probabilistic} represents samples as gaussian distributions in the embedding space and trains their framework using a contrastive loss between the sample distributions computed by Monte-Carlo sampling. Chun~\cite{chun2023improved} later introduced PCME++, which improves PCME by using a closed-form distance formulation, eliminating the need for sampling. We adopt PCME++ to learn a probabilistic embedding space for audio, its textual description, and multi-scale satellite imagery.

\section{Methods}

\begin{figure*}
    \centering
    \includegraphics[width=\textwidth]{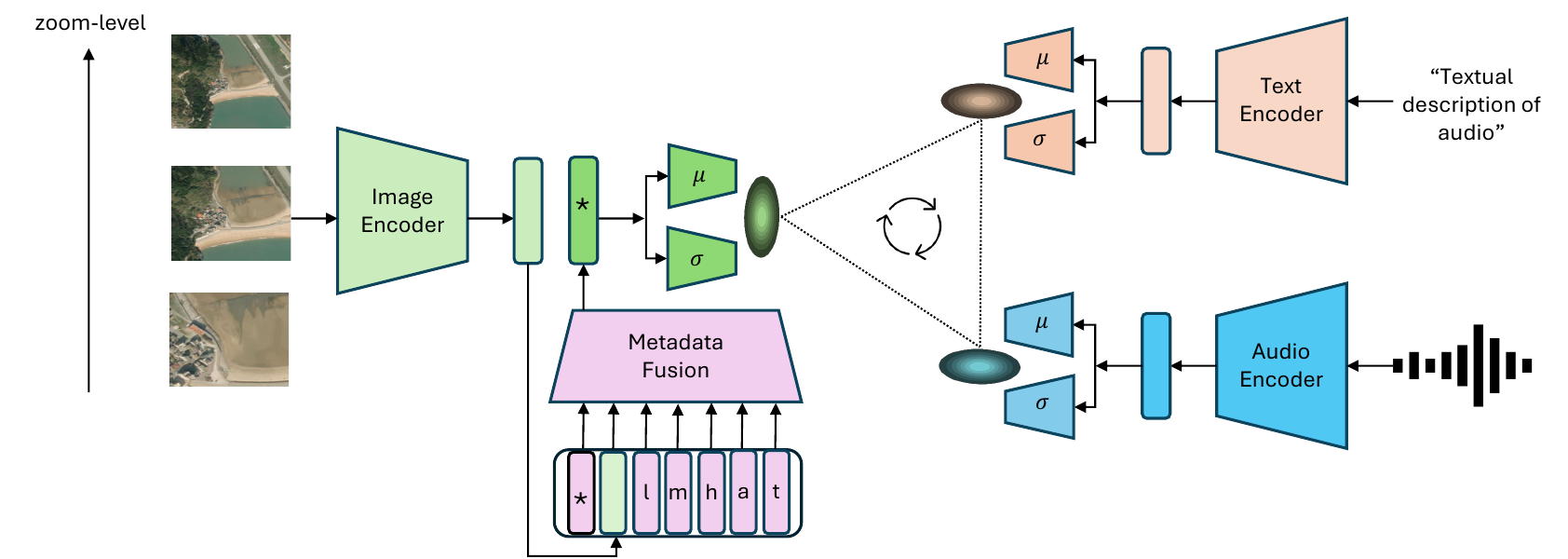}
    \caption{Our proposed framework, Probabilistic Soundscape Mapping (PSM), combines image, audio, and text encoders to learn a probabilistic joint representation space. Metadata, including geolocation (l), month (m), hour (h), audio-source (a), and caption-source (t), is encoded separately and fused with image embeddings using a transformer-based metadata fusion module. For each encoder, $\mu$ and $\sigma$ heads yield probabilistic embeddings, which are used to compute probabilistic contrastive loss.}
    \label{fig:teaser}
\end{figure*}

\label{sec:method}
\subsection{Dataset Creation}

Prior work in zero-shot soundscape mapping \cite{khanal2023soundscape} has utilized the \textit{SoundingEarth} dataset \cite{heidler2023self}, which contains approximately $50$k geotagged audios paired with corresponding satellite imagery. To facilitate research on training large-scale models with a rich representation space for soundscape mapping, we have expanded the size of the dataset $6$-fold by creating a dataset containing \num{309019} geo-tagged audios from four different sources: \textit{iNaturalist}~\cite{iNaturalist}, \textit{yfcc-video}~\cite{thomee2016yfcc100m}, \textit{aporee}~\cite{aporee}, and \textit{freesound}~\cite{Freesound}, each contributing \num{114603}, \num{96452}, \num{49284}, and \num{48680} samples respectively. We pair these geotagged audios with their corresponding \textit{Sentinel-2-cloudless} imagery \cite{eoxmaps} with $10$m GSD and \textit{Bing} imagery with $0.6$m GSD. 

In the prior work, GeoCLAP~\citep{khanal2023soundscape}, samples were randomly split between train/validation/test sets for training and evaluating their models. We observed that such a data split strategy leads to the issue of data leakage where evaluation data samples come from the same set of locations present in the training set, preventing the evaluation of the generalizability of a model to truly unseen locations. To address this, we divide the world into $1$\textdegree~$\times$~$1$\textdegree~non-overlapping cells where each cell containing some samples is randomly assigned to either train/validation/test set. Our dataset contains \num{294019}/\num{5000}/\num{10000} samples in the train/validation/test sets. We also employ our split strategy on the \textit{SoundingEarth} dataset with a cell size of $10km$~$\times$~$10km$. This strategy resulted in  \num{41469}/\num{3242}/\num{5801} samples in train/validation/test sets. Details of our dataset and split strategy are in the supplemental material.   

\subsection{Approach}
This section describes our framework (PSM) for learning a metadata-aware, probabilistic, and tri-modal embedding space for multi-scale zero-shot soundscape mapping.

Figure~\ref{fig:teaser} presents an overview of the PSM framework, which comprises an image encoder, metadata fusion module, text encoder, and audio encoder. The scale-aware image encoder converts multi-scale satellite imagery into a $d$-dimensional representation. The transformer-based metadata fusion module integrates metadata (including location, month, time, audio source, and text source) with the image representation, generating a metadata-aware probabilistic image representation. Other modality-specific encoders produce probabilistic embeddings for text and audio. PSM aims to map tuples of satellite imagery, audio, and text into a shared probabilistic representation space.

Given a geotagged audio~$X_{k}^{a}$, textual description of the audio~$X_{k}^{t}$, and a satellite image at a given location viewed at a zoom level~$l$ (an integer between 1 and some maximum zoom level $L$)~$X_{k,l}^{i}$,  ($X_{k}^{a}$,$X_{k}^{t}$,$X_{k,l}^{i}$) is the $k$-th audio-text-image triplet. PSM is trained over the aggregation of all available triplets. 

We use modality-specific transformer-based encoders followed by their respective linear projection modules to obtain representations ($h_{k}^{a}$,$h_{k}^{t}$,$h_{k,l}^{i}$) with same dimension $d$, 
\begin{equation}\label{eq:3}
h_{k}^{a} = g_{audio}(f_{audio}(X_{k}^{a}))
\end{equation}
\begin{equation}\label{eq:4}
h_{k}^{t} = g_{text}(f_{text}(X_{k}^{t}))
\end{equation}
\begin{equation}\label{eq:5}
h_{k}^{i} = g_{image}(f_{image}(X_{k}^{i}, l_{k}))
\end{equation}
where $(f_{audio}, g_{audio})$, $(f_{text}, g_{text})$, $(f_{image}, g_{image})$  are (encoder, projection-module) pairs producing $d$ dimensional embeddings: $h_{k}^{a}$, $h_{k}^{t}$, and $h_{k}^{i}$, for audio, text, and satellite image with zoom-level $l_{k}$ respectively. 

We use GSDPE~\cite{reed2023scale} to encode the position and scale of each patch of satellite imagery at zoom-level ($l$) to learn scale-aware representations of multi-scale satellite imagery,
\begin{equation}\label{eq:1}
v_{l,x}(pos,2i) = sin(\frac{g*l}{G})\frac{pos}{10000^{\frac{2i}{d}}}
\end{equation}
\begin{equation}\label{eq:2}
v_{l,y}(pos,2i+1) = cos(\frac{g*l}{G})\frac{pos}{10000^{\frac{2i}{d}}}
\end{equation} where $pos$ is the position of the image patch along the given axis ($x$ or $y$), $i$ is the feature dimension index, $l$ is the zoom-level of the image, $g$ is the GSD of the image, and $G$ is the reference GSD. 

As discussed before, we are interested in learning metadata-aware representation space. Therefore, we fuse four different components of metadata (geolocation, month, hour, audio-source, caption-source) with the satellite image embedding ($h_{k}^{i}$) and obtain a metadata-conditioned image embedding ($h_{k}^{i'}$),
\begin{equation}\label{eq:6}
h_{k}^{i'} = g_{meta}(h_{k}^{i}, metadata)
\end{equation}
where $g_{meta}$ is a transformer based metadata fusion module of our framework, $h_{k}^{i'}$ is the embedding corresponding to the learnable special token (*) fed into $g_{meta}$.

To learn a probabilistic embedding space, we define the embedding of a given modality ($r$) as a normally distributed random variable, $Z_{r}\sim N(\mu_{r}, \sigma_{r})$. We employ a closed-form probabilistic contrastive loss ~\cite {chun2023improved} between all three pairs of embeddings.  For any two modalities $p$ and $q$, the closed-form sampled distance (CSD) as defined in PCME++ ~\cite{chun2023improved} is as follows:
\begin{equation}\label{eq:7}
d(Z_{p},Z_{q})  
= \lVert \mu_{p} -\mu_{q}\rVert ^{2}_{2} + \lVert \sigma_{p}^{2} +\sigma_{q}^{2}\rVert _{1}.
\end{equation}
In our implementation, we pass our modality-specific representations, $h_{k}^{a}$, $h_{k}^{t}$, and $h_{k}^{i'}$, through heads for $\mu$ and $\log(\sigma^{2})$ of the gaussian distribution representing our samples.

This distance is used to compute matching loss between the pairs as shown in Equation \ref{eq:8}. For positive (certain) matches, to reduce distance, $\mu$ embeddings are pulled closer together while minimizing $\sigma$ embeddings in $\lVert \sigma_{p}^{2} +\sigma_{q}^{2}\rVert _{1}$. Similarly, for negative (uncertain) match, to increase the distance, $\mu$ embeddings are pushed farther while maximizing $\sigma$ embeddings in $\lVert \sigma_{p}^{2} +\sigma_{q}^{2}\rVert _{1}$. 

Based on the distance function defined in Equation \ref{eq:7}, we can define the probabilistic matching objective function as follows:
\begin{equation}\label{eq:8}
\begin{split}
\mathcal{L}_{m} = -[w_{pq}\log(\text{sigmoid}(-a.d(Z_p,Z_q)+b))+ \\
(1-w_{pq})\log(\text{sigmoid}(a.d(Z_p,Z_q)-b))]
\end{split}
\end{equation}
where $w_{pq} \in \{0,1\}$ is the matching indicator between $p$ and $q$. $a$ and $b$ are learnable scalar parameters. $\mathcal{L}_{m}$ ($\mathcal{L}_{match}$) is computed for all sample pairs in the mini-batch. 

Soundscape mapping is inherently a one-to-many matching problem. Given a satellite image at a location, there may be multiple sounds that are likely to be heard there. Therefore, if we were to simply assign $w_{pq}$ as $0$ or $1$ for our dataset's negative and positive matches, we would lose the opportunity to learn from the potentially numerous false negatives. Therefore, we adopt a similar strategy of learning from pseudo-positives, as formulated by Chun~\cite{chun2023improved}. In this approach, for a positive match ($p$,$q$), we consider $q'$ as a pseudo-positive match (\textit{pseudo-m}) with $q$ if $d(Z_{p}, Z_{q'}) \leq d(Z_{p}, Z_{q})$. Finally, the objective function for a pair of modalities ($p,q$) is as follows:
\begin{equation}\label{eq:9}
\mathcal{L}_{p,q} = \mathcal{L}_{m} + \alpha\mathcal{L}_{pseudo-m} + \beta\mathcal{L}_{VIB}
\end{equation}
where $\alpha$ and $\beta$ control for the contribution of pseudo-match loss and Variational Information Bottleneck (VIB) loss~\cite{alemi2019deep}, respectively. We use $\mathcal{L}_{VIB}$ as a regularizer to reduce overfitting, preventing the collapse of $\sigma$~to 0.

To learn a tri-modal embedding space for zero-shot soundscape mapping, using Equation~\ref{eq:9}, we separately compute loss for all three pairs of modalities: audio-text ($a,t$), audio-image ($a,i$), and image-text($i,t$). Finally, the overall objective function to train PSM is as follows:
\begin{equation}\label{eq:10}
\mathcal{L} = \mathcal{L}_{a,t} + \mathcal{L}_{a,i} + \mathcal{L}_{i,t}.
\end{equation}

\begin{table*}[]
\centering
\begin{tabular}{@{}lccccc|cccc@{}}
\toprule
method & loss & meta/train & text/eval & meta/eval & ZL & I2A R@10\% & I2A MdR & A2I R@10\% & A2I MdR\\ \midrule
GeoCLAP & infoNCE & \xmark & \xmark & \xmark                                   & 1 & 0.399 & 1500 & 0.403 & 1464   \\
GeoCLAP & infoNCE & \xmark & \cmark  & \xmark                                   & 1 & 0.577 & 712  & 0.468 & 1141   \\ \midrule
ours     & infoNCE & \cmark & \cmark  & \cmark & 1 & 0.709 & 462  & 0.871 & 241  \\
ours     & PCME++  & \xmark & \xmark & \xmark                                   & 1 & 0.423 & 1401 & 0.428 & 1344 \\
ours     & PCME++  & \cmark & \xmark & \cmark & 1 & 0.828 & 261  & 0.829 & 248  \\
ours     & PCME++  & \cmark & \cmark  & \cmark & 1 & \textbf{0.901} & \textbf{113}  & \textbf{0.943} & \textbf{100}  \\ \midrule

GeoCLAP & infoNCE & \xmark & \xmark & \xmark                                   & 3 & 0.408 & 1441 & 0.420  & 1389 \\
GeoCLAP & infoNCE & \xmark & \cmark  & \xmark                                   & 3 & 0.577 & 707  & 0.483 & 1056 \\ \midrule
ours     & infoNCE & \cmark & \cmark  & \cmark & 3 & 0.708 & 462  & 0.875 & 235  \\
ours     & PCME++  & \xmark & \xmark & \xmark                                   & 3 & 0.440 & 1302 & 0.443 & 1266 \\
ours     & PCME++  & \cmark & \xmark & \cmark & 3 & 0.827 & 266  & 0.832 & 250  \\
ours     & PCME++  & \cmark & \cmark  & \cmark & 3 & \textbf{0.900} & \textbf{114}  & \textbf{0.945} & \textbf{102}  \\ \midrule

GeoCLAP & infoNCE &  \xmark & \xmark & \xmark                                   & 5 & 0.409 & 1428 & 0.421 & 1373 \\
GeoCLAP & infoNCE &  \xmark & \cmark  & \xmark                                   & 5 & 0.581 & 698  & 0.489 & 1036 \\ \midrule
ours     & infoNCE & \cmark & \cmark  & \cmark & 5 & 0.709 & 461  & 0.875 & 238  \\
ours     & PCME++  & \xmark & \xmark & \xmark                                   & 5 & 0.440 & 1302 & 0.448 & 1279 \\
ours     & PCME++  & \cmark & \xmark & \cmark & 5 & 0.821 & 281  & 0.826 & 261  \\
ours     & PCME++  & \cmark & \cmark  & \cmark & 5 & \textbf{0.896} & \textbf{115}  & \textbf{0.941} & \textbf{107}  \\ \bottomrule
\end{tabular}
\caption{Experimental results for models trained on the GeoSound dataset with satellite imagery from \textit{Bing}.}
\label{tab:bing}
\end{table*}
\begin{table*}[]
\centering
\begin{tabular}{@{}lccccc|cccc@{}}
\toprule
method & loss & meta/train & text/eval & meta/eval & ZL & I2A R@10\% & I2A MdR & A2I R@10\% & A2I MdR\\ \midrule
GeoCLAP & infoNCE & \xmark & \xmark & \xmark                                   & 1 & 0.459 & 1179 & 0.465 & 1141 \\
GeoCLAP & infoNCE & \xmark & \cmark  & \xmark                                   & 1 & 0.546 & 827  & 0.553 & 804  \\ \midrule
ours     & infoNCE & \cmark & \cmark  & \cmark & 1 & 0.722 & 497  & 0.860  & 247  \\
ours     & PCME++  & \xmark &  \xmark & \xmark                                   & 1 & 0.474 & 1101 & 0.485 & 1061 \\
ours     & PCME++  & \cmark & \xmark & \cmark & 1 & 0.802 & 294  & 0.804 & 283  \\
ours     & PCME++  & \cmark & \cmark  & \cmark & 1 & \textbf{0.872} & \textbf{142}  & \textbf{0.940}  & \textbf{104}  \\ \midrule

GeoCLAP & infoNCE & \xmark & \xmark & \xmark                                   & 3 & 0.454 & 1200 & 0.456 & 1197 \\
GeoCLAP & infoNCE & \xmark & \cmark  & \xmark                                   & 3 & 0.542 & 840  & 0.555 & 790  \\ \midrule
ours     & infoNCE & \cmark & \cmark  & \cmark & 3 & 0.722 & 491  & 0.856 & 248  \\
ours     & PCME++  & \xmark & \xmark & \xmark                                   & 3 & 0.479 & 1086 & 0.487 & 1042 \\
ours     & PCME++  & \cmark & \xmark & \cmark & 3 & 0.795 & 306  & 0.800   & 290  \\
ours     & PCME++  & \cmark &  \cmark  & \cmark & 3 & \textbf{0.870}  & \textbf{150}  & \textbf{0.940}  & \textbf{104}  \\ \midrule

GeoCLAP & infoNCE & \xmark & \xmark & \xmark                                   & 5 & 0.458 & 1194 & 0.457 & 1184 \\
GeoCLAP & infoNCE & \xmark & \cmark  & \xmark                                   & 5 & 0.542 & 835  & 0.554 & 791  \\ \midrule
ours     & infoNCE & \cmark & \cmark  & \cmark & 5 & 0.719 & 497  & 0.852 & 252  \\
ours     & PCME++  & \xmark & \xmark & \xmark                                   & 5 & 0.459 & 1172 & 0.465 & 1138 \\
ours     & PCME++  & \cmark & \xmark & \cmark & 5 & 0.794 & 316  & 0.794 & 299  \\
ours     & PCME++  & \cmark & \cmark  & \cmark & 5 & \textbf{0.868} & \textbf{156}  & \textbf{0.935} & \textbf{109}  \\ \bottomrule
\end{tabular}
\caption{Experimental results for models trained on the GeoSound dataset with satellite imagery from \textit{Sentinel-2}.}
\label{tab:sentinel}
\end{table*}
\begin{table*}[]
\centering
\begin{tabular}{@{}lcccc|cccc@{}}
\toprule
method & loss & meta/train & text/eval & meta/eval & I2A R@10\% & I2A MdR & A2I R@10\% & A2I MdR\\ \midrule
GeoCLAP & infoNCE & \xmark & \xmark & \xmark                          & 0.454 & 667 & 0.449 & 694 \\
GeoCLAP & infoNCE & \xmark & \cmark  & \xmark                          & 0.523 & 533 & 0.470 & 641 \\ \midrule
ours     & infoNCE & \cmark & \cmark  & \cmark & 0.519 & 548 & 0.491 & 596 \\
ours     & PCME++  & \xmark & \xmark & \xmark                          & 0.514 & 547 & 0.518 & 543 \\
ours     & PCME++  & \cmark & \xmark & \cmark & 0.563 & 454 & 0.569 & 447 \\
ours     & PCME++  & \cmark & \cmark  & \cmark & \textbf{0.690} & \textbf{264} & \textbf{0.608} & \textbf{371} \\ \bottomrule
\end{tabular}
\caption{Experimental results for models trained on the SoundingEarth dataset.}
\label{tab:soundingearth-main}
\end{table*}

\section{Experimental Details}
\label{sec:experimental_details}
\textbf{Audio/Text Processing:}
We use pre-trained models for the audio and text modalities and their respective input processing pipelines hosted on \texttt{HuggingFace}. Specifically, for audio, we extract the audio spectrogram using the \texttt{ClapProcessor} 
wrapper for the pre-trained CLAP~\cite{laionclap2023} model \texttt{clap-htsat-fused} 
with default parameters: \texttt{feature\_size=64}, \texttt{sampling\_rate=48000}, \texttt{hop\_length=480}, \texttt{fft\_window\_size=1024}. CLAP uses a feature fusion strategy~\cite{laionclap2023} to pre-process variable length 
sounds by extracting a spectrogram of randomly selected 3 $d$-second audio slices and the spectrogram of the whole audio down-sampled to 10s. We choose $d=$10s in our experiments. Apart from the text present in the metadata, we also obtain a textual description of audio from a recent SOTA audio captioning model, Qwen-sound~\cite{Qwen-Audio}, and use the captioning model's output only if it passes CLAP-score~\cite{laionclap2023} based quality check. For the textual descriptions of audio in our data, we adopt the similar text processing as performed by CLAP~\cite{laionclap2023} and tokenize our text using \texttt{RobertaTokenizer} with \texttt{max\_length=128}.

\textbf{Satellite image processing:} Our framework is trained with satellite images at different zoom levels $l\in \{1, 3, 5\}$. To obtain this data, we first downloaded a large tile of images with size $(Lh)\times(Lw)$. We obtained high-resolution $0.6$m GSD imagery with a tile size of $1500 \times 1500$ from \textit{Bing} and low-resolution 10m GSD imagery with a tile size of $1280 \times 1280$ from \textit{Sentinel-2-cloudless}. To get an image at zoom-level $l$, we center crop from the original tile with a crop size of $(lh)\times(lw)$ and then resize it to an $h \times w$ image, where $(h,w)$ is $(256,256)$ for \textit{Sentinel-2} imagery and $(300,300)$ for \textit{Bing} imagery. This way, we can simulate the effect of change in coverage area as the zoom-level changes while effectively keeping constant input image size for training. During training, we randomly sample $l$ from a set \{1, 3, 5\} for each image instance. Then, for the zoom-transformed image, we perform \textit{RandomResizedCrop} with parameters: \texttt{\{input\_size=224, scale=(0.2, 1.0)\}} followed by a \textit{RandomHorizontalFlip} while only extracting a $224 \times 224$ center crop of the image at the desired zoom-level $l$ for evaluation.

\textbf{Metadata Fusion:} To fuse metadata into our framework, we first separately project the metadata components into $512$-dimensional space using linear layers and concatenate them with the satellite image embedding from the image encoder and a learnable special token. Finally, the set of tokens is fed into a lightweight transformer-based module containing only $3$ layers. The output of this module is further passed through heads for $\mu$ and $\log(\sigma^{2})$ of the Gaussian distribution representing metadata-conditioned image embeddings. To avoid overfitting to the metadata, we independently drop each metadata component at the rate of $0.5$ during training.

\begin{table*}[]
\centering
\begin{tabular}{@{}lccccc|cccc@{}}
\toprule
imagery  & latlong & month & time & audio source & text source & I2A R@10\% & I2A MdR & A2I R@10\% & A2I MdR \\ \midrule
Sentinel-2 & \cmark  & \xmark & \xmark & \xmark & \xmark & 0.512 & 946  & 0.516 & 923  \\
Sentinel-2 & \xmark & \cmark  & \xmark & \xmark & \xmark & 0.501 & 988  & 0.511 & 941  \\
Sentinel-2 & \xmark & \xmark & \cmark  & \xmark & \xmark & 0.548 & 825  & 0.574 & 717  \\
Sentinel-2 & \xmark & \xmark & \xmark & \cmark  & \xmark & \textbf{0.749} & \textbf{407}  & \textbf{0.757} & \textbf{389}  \\
Sentinel-2 & \xmark & \xmark & \xmark & \xmark & \cmark  & 0.483 & 1080 & 0.492 & 1022 \\ \midrule
Bing  & \cmark  & \xmark & \xmark & \xmark & \xmark & 0.539 & 822  & 0.557 & 764  \\
Bing  & \xmark & \cmark  & \xmark & \xmark & \xmark & 0.464 & 1153 & 0.485 & 1068 \\
Bing  & \xmark & \xmark & \cmark  & \xmark & \xmark & 0.516 & 937  & 0.547 & 823  \\
Bing  & \xmark & \xmark & \xmark & \cmark  & \xmark & \textbf{0.722} & \textbf{469}  & \textbf{0.733} & \textbf{447}  \\
Bing  & \xmark & \xmark & \xmark & \xmark & \cmark  & 0.448 & 1250 & 0.466 & 1140 \\ \bottomrule
\end{tabular}
\caption{Metadata ablation to evaluate the impact of individual metadata components on the best model's performance. The best model for each imagery type was trained with all metadata types and can handle multiple zoom levels. The results presented are specific to satellite imagery at zoom level 1, with no text embeddings added to the query embeddings during inference.}
\label{tab:ablation-meta}
\end{table*}

\textbf{Training:}  We initialize encoders from released weights of pre-trained models, SatMAE~\cite{satmae2022} for satellite imagery and CLAP~\cite{laionclap2023} for audio and text. We chose $d$, the dimensionality of our embeddings, to be $512$. For regularization, we set the weight decay to $0.2$. Our training $batch\_size$ was $128$. We use \texttt{Adam} as our optimizer, with the initial learning rate set to $5e-5$. To schedule the learning rate, we use cosine annealing with warm-up iterations of $5k$ for experiments with \textit{GeoSound} and $2k$ for experiments with \textit{SoundingEarth}.

\textbf{Baseline:} We use GeoCLAP~\cite{khanal2023soundscape}, a SOTA zero-shot soundscape mapping model, as a baseline for evaluation. GeoCLAP is contrastively trained using the \textit{infoNCE}~\cite{infonce} loss between three modality pairs: image-audio, audio-text, and image-text.

\textbf{Metrics:} We evaluate on two datasets: \textit{GeoSound}, and \textit{SoundingEarth}. We use Recall@10\% and the Median Rank of the ground truth as our evaluation metrics.  Recall@10\% (R@10\%) is defined as the proportion of queries for which the ground-truth match is found within the top 10\% of the returned ranked retrieval list. Median Rank (MdR) is defined as the median of the positions at which the ground-truth matches appear in the ranked retrieval list. We denote image-to-audio as I2A and audio-to-image as A2I throughout the paper. To assess the effectiveness of text embeddings in cross-modal retrieval between satellite images and audio, we also evaluate an experimental setting where, during inference, we add the corresponding text embedding to the query embedding during retrieval.

\section{Results}
\label{sec:results}

\subsection{Cross-Modal Retrieval with Bing}

Table~\ref{tab:bing} presents our retrieval evaluation of {\name} trained on the \textit{GeoSound} dataset using \textit{Bing} satellite imagery. Our approach outperforms the state-of-the-art baseline~\cite{khanal2023soundscape} for cross-modal retrieval between satellite imagery and audio, and vice versa. SatMAE \cite{satmae2022} with GSDPE is utilized to encode the zoom level of the satellite imagery for both the baseline and our models. This enables our satellite image encoder to remain invariant to zoom-level changes, achieving consistent performance across all zoom levels. We observe that learning a probabilistic embedding space using PCME++ loss alone enhances the baseline performance from 0.399 to 0.423, 0.408 to 0.440, and 0.409 to 0.440 for zoom levels 1, 3, and 5, respectively. In addition to the objective function, we also experimented with the inclusion of metadata during training and inference. As anticipated, the model's performance, when trained and evaluated with both text and metadata, is notably improved, enhancing image-to-audio  R@10\% from the baseline score of 0.577 to 0.901, 0.577 to 0.900, and 0.581 to 0.896 for zoom levels 1, 3, and 5, respectively. A similar trend is observed for audio-to-image retrieval.

\begin{figure*}[h!]
\centering
\includegraphics[width=\linewidth]{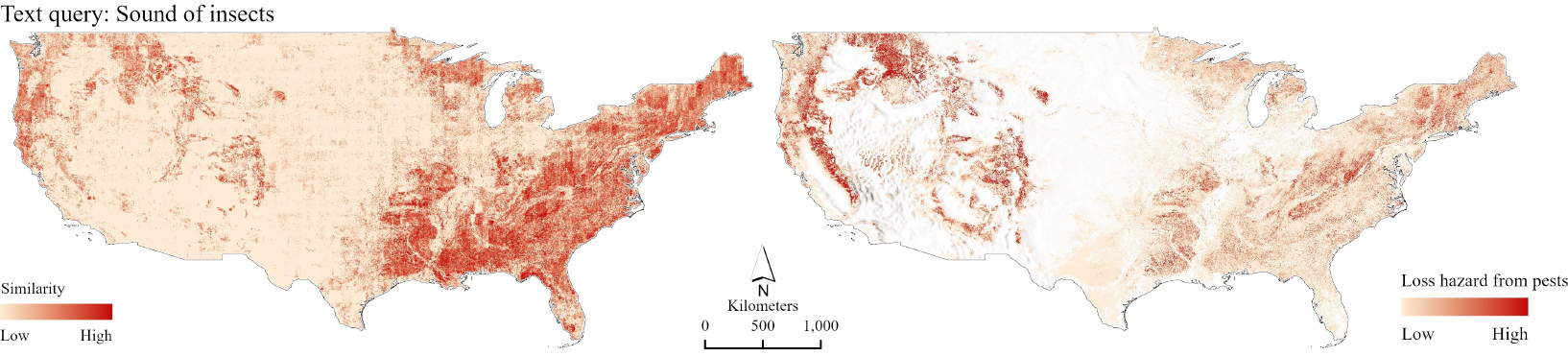}
\caption{Soundscape Map of the USA for a textual query \textit{Sound of insects}, compared with a reference map \cite{NIDRM} indicating the risk of pest-related hazard.}
\label{fig:insects_soundscapes}
\end{figure*}

\begin{figure}[h!]
\centering
\includegraphics[width=\columnwidth]{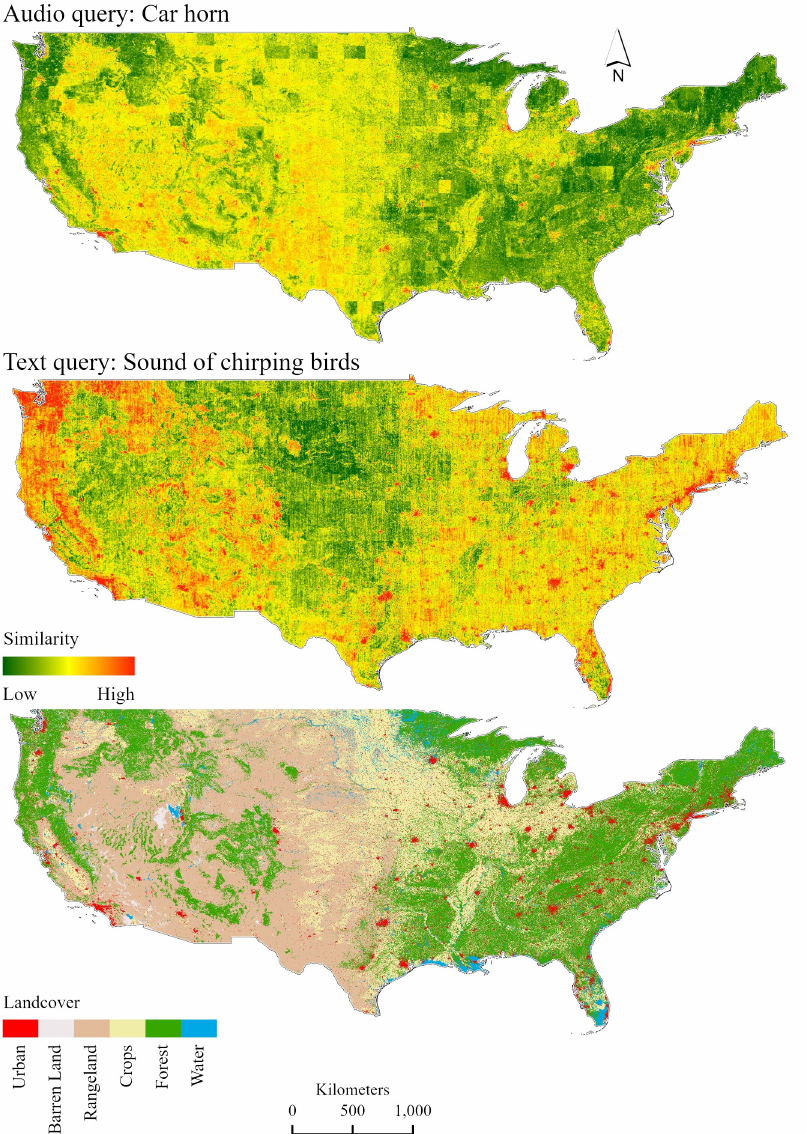}
\caption{Two soundscape maps of the continental United States, generated using different query types, with a land cover map \cite{landcover} for reference.}
\label{fig:usa_soundscapes}
\end{figure}
 
\begin{figure*}[ht!]
        \centering
        \includegraphics[width=\linewidth]{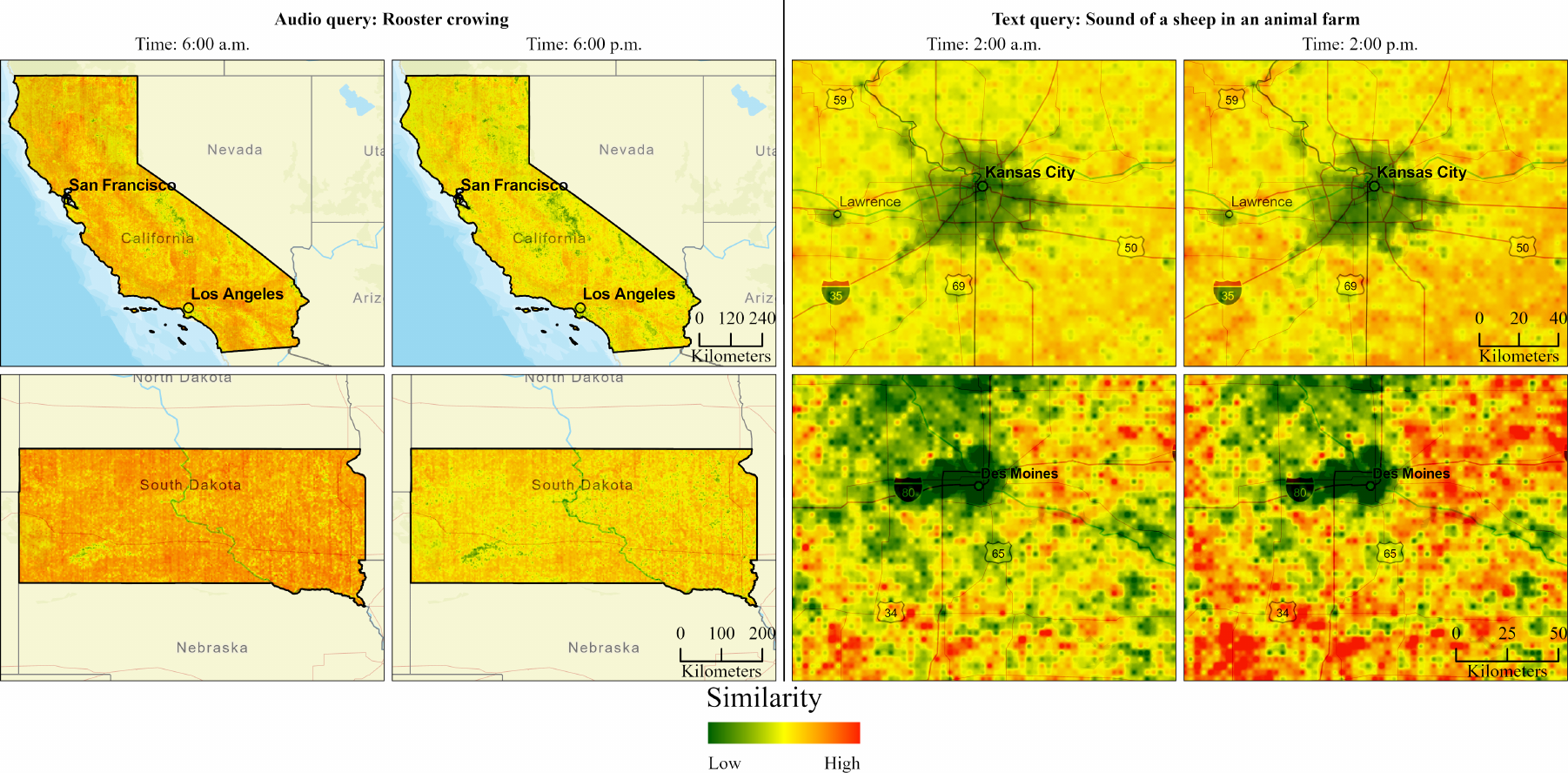}
        \caption{Temporally dynamic soundscape maps created by querying our model for different geographic areas.}
        \label{fig:dynamic_maps}
    \end{figure*}

\subsection{Cross-Modal Retrieval with Sentinel-2}
Table~\ref{tab:sentinel} presents the evaluation results of {\name} trained on the \textit{GeoSound} dataset using \textit{Sentinel-2} satellite imagery. Similar to results with \textit{Bing} imagery, we observe consistent performance across various zoom levels, indicating the robustness of our framework in extracting valuable information irrespective of the coverage area of input satellite imagery. By employing PCME++ loss in training our framework, we note an enhancement in the baseline performance from 0.459 to 0.474 for zoom level 1. Overall, {\name} trained with \textit{Sentinel-2} imagery and metadata, and evaluated using both metadata and text during inference, significantly improved the baseline score from 0.546 to 0.872, 0.542 to 0.870, and 0.542 to 0.868 for zoom levels 1, 3, and 5, respectively. A similar trend is observed for audio-to-image retrieval. Moreover, the high performance of {\name} on \textit{Sentinel-2} imagery at zoom level 5 enables the efficient creation of large-scale soundscape maps using freely available \textit{Sentinel-2} imagery while requiring fewer images to download.

\subsection{Cross-Modal Retrieval on SoundingEarth}
Table~\ref{tab:soundingearth-main} presents the evaluation results of {\name} trained on the \textit{SoundingEarth} dataset~\cite{heidler2023self} with its original $0.2$m GSD \textit{GoogleEarth} imagery. For the \textit{SoundingEarth} dataset, our models are exclusively trained and evaluated on zoom level 1. Similar to the performance observed on the \textit{GeoSound} dataset, we witness gain in performance with our approach of learning a metadata-aware probabilistic embedding space. Specifically, by training with the PCME++ objective instead of the \textit{infoNCE} loss, we note an improvement in the score from 0.454 to 0.514. This performance further elevates to 0.563 when metadata is incorporated and reaches 0.690 when both metadata and text are utilized during inference. We observe similar trends for audio-to-image retrieval as well.

\label{sec:discussion}

\subsection{Effect of Metadata}
Our experimental results reveal a significant enhancement in the model's performance when metadata is integrated into both training and inference. For comparison, as illustrated in Table \ref{tab:bing}, {\name} trained with \textit{Bing} imagery without any metadata achieved an I2A R@10\% of 0.423, whereas with all metadata included, it reached 0.828. A similar trend is seen for experiments with \textit{Sentinel-2} imagery. {\name} is designed such that individual metadata components are independently masked out with a rate of 0.5. Therefore, during inference, we can evaluate {\name} by dropping any combination of metadata components.

In Table \ref{tab:ablation-meta}, we present the ablation of different metadata components to evaluate the impact of individual metadata components in {\name}'s learning framework. 
We perform this ablation on our best-performing models trained on the \textit{GeoSound} dataset with \textit{Sentinel-2} and \textit{Bing} imagery. The results in Table \ref{tab:ablation-meta} exclude the use of text during cross-modal retrieval and utilize satellite imagery at zoom level 1 during inference.
These results highlight two major findings. First, all of the metadata components contribute to the overall improvement of {\name}'s performance. Second, among all of the metadata components, audio-source had the most significant impact. This suggests that biases in different audio hosting platforms are encoded into the learning framework, improving cross-modal retrieval and enabling soundscape maps conditioned on the expected audio type from specific platforms.
    
\subsection{Generating Country-Level Soundscape Maps}
We demonstrate {\name}'s capability to generate large-scale soundscape maps using audio and text queries. We acquired 0.6 m GSD $1500 \times 1500$ image tiles encompassing the entire USA from \textit{Bing}. 


Using our top-performing model's image encoder, we pre-compute embeddings for each image at zoom level $1$. During inference, these are combined with desired metadata embeddings via the model's metadata fusion module to obtain metadata-conditioned probabilistic embeddings for the entire region.


We use modality-specific encoders to obtain probabilistic embeddings for audio or text queries. To compute similarity scores between image embeddings and query embeddings, we utilize Equation ~\ref{eq:7} as detailed in our paper. These scores are then used to produce large-scale soundscape maps, as shown in Figures \ref{fig:insects_soundscapes} and \ref{fig:usa_soundscapes}.

\section{Discussion}

Figure \ref{fig:insects_soundscapes} depicts a soundscape map generated for the textual query \textit{``Sound of insects''}, accompanied by the following metadata: \texttt{\{audio source: iNaturalist, month: May, time: 8 pm\}}. Notably, this soundscape map exhibits a strong correlation with an available reference map \cite{NIDRM}, which shows potential pest hazards across the continental United States. Figure \ref{fig:usa_soundscapes} showcases two soundscape maps: one for an audio query of \textit{car horn} with the metadata \texttt{\{audio source: yfcc, month: May, time: 10 am\}}, and another for a textual query \textit{``Sound of chirping birds.''} with metadata: \texttt{\{audio source: iNaturalist, month: May, time: 10 am\}}. Both maps can be compared with a land cover map~\cite{landcover}. As expected, for the car horn query, higher activation is observed in most major US cities, while for chirping birds, increased activation is observed around both urban areas and forests.

We also note that the soundscape of any geographic region evolves predictably over a day. Therefore, the hour of the day is one of the important metadata components fused into our framework. In addition to contributing to increased performance, temporal understanding fused into our embedding space allows us to create temporally dynamic soundscape maps across any geographic region, as demonstrated in Figure~\ref{fig:dynamic_maps}. The similarity scores used for these soundscape maps were normalized consistently for a region across time. We display state-level temporally dynamic soundscape maps for an audio query: \textit{Rooster crowing} with metadata: \texttt{\{audio source: aporee, month: May, time: 6 am\}} vs. \texttt{\{audio source: aporee, month: May, time: 6 pm\}}. We observe that for both states, higher activation for the \textit{rooster crowing} audio query is seen on the soundscape map at 6 am. We also showcase city-level temporally dynamic soundscape maps for a text query \textit{``Sound of a sheep in an animal farm.''}. We can observe that for areas around both cities, Kansas City and Des Moines, very low activation is present. Additionally, higher activation is observed at 2 pm than at 2 am, which is expected. These demonstrations highlight the ability of our model to create semantically meaningful and temporally consistent soundscape maps across any geographic regions of interest.
\section{Conclusion}
\label{sec:conclusion}

Our work introduces a framework for learning probabilistic tri-modal embeddings for the task of multi-scale zero-shot soundscape mapping. To advance research in this direction, we have developed a new large-scale dataset that pairs geotagged audio with high and low-resolution satellite imagery. By utilizing a probabilistic tri-modal embedding space, our method surpasses the state-of-the-art while also providing uncertainty estimates for each sample. Furthermore, we have designed our framework to be metadata-aware, resulting in a significant improvement in cross-modal retrieval performance. Additionally, it enables the creation of dynamic soundscape maps conditioned on different types of metadata. The combination of enhanced mapping performance, uncertainty estimation, and a comprehensive understanding of spatial and temporal dynamics positions our framework as an effective solution for zero-shot multi-scale soundscape mapping.


\bibliographystyle{ACM-Reference-Format}
\bibliography{main}
\newpage
\appendix 
\section{Dataset Creation}
\label{sec:data}
\begin{figure}[h!]
    \centering
    \includegraphics[width=\linewidth]{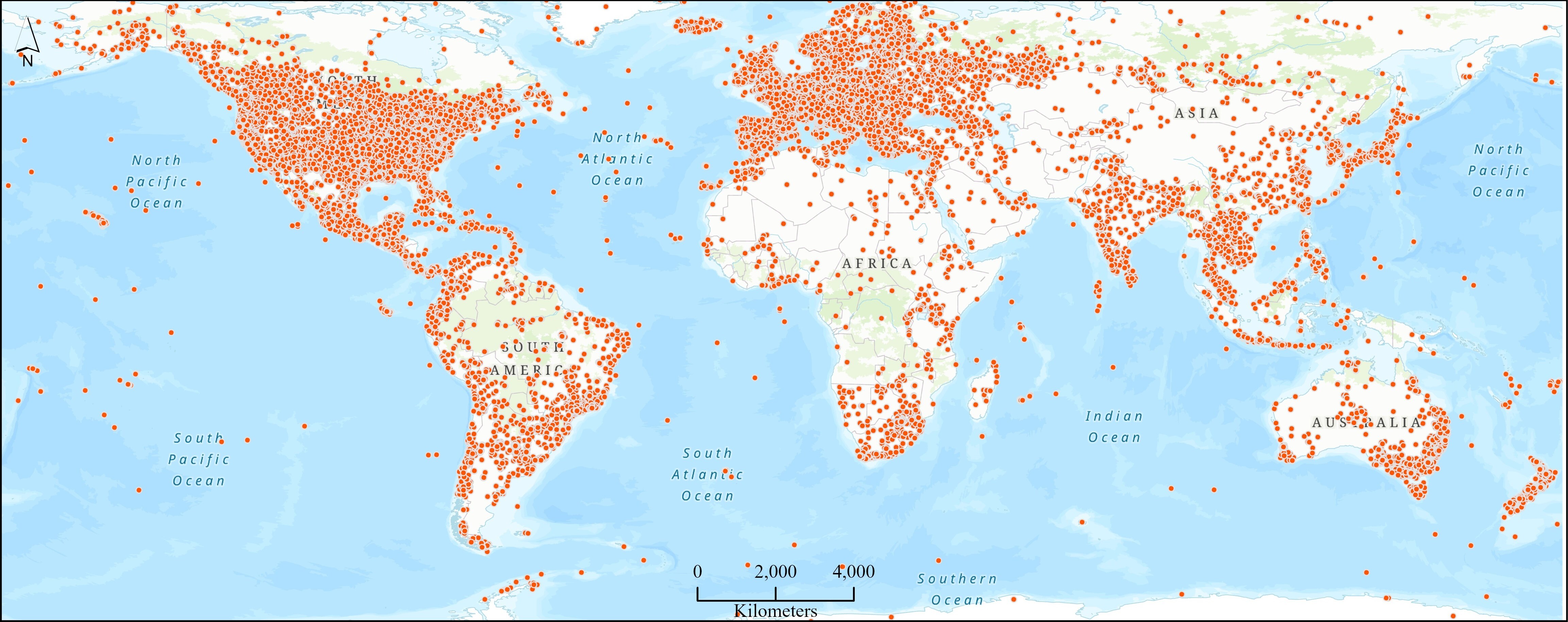}
    \caption{Distribution of samples in the \textit{GeoSound} dataset.}
    \label{fig:data_map}
\end{figure}
We have created a new large-scale dataset (\textit{GeoSound}) suitable for the task of zero-shot soundscape mapping, effectively increasing the size of available dataset~\citep{heidler2023self} by more than $6$-fold. To achieve this, we collected geotagged audios along with associated metadata (textual description, geolocation, time) from four different audio sources: \textit{iNaturalist}~\cite{iNaturalist}, \textit{YFCC100M}~\cite{thomee2016yfcc100m}, \textit{Radio Aporee}~\cite{aporee}, and \textit{Freesound}~\cite{Freesound}. For each of the audio samples in our dataset, we downloaded $1500 \times 1500$ high-resolution ($0.6$m GSD) imagery from \textit{Bing} and $1280 \times 1280$ low-resolution ($10$m GSD) \textit{Sentinel-2 Cloudless} imagery from \textit{EOX::Maps}~\cite{eoxmaps}.
Figure~\ref{fig:data_map} illustrates the geospatial distribution of data samples in the \textit{GeoSound} dataset worldwide.

\subsection{Audio Sources}
\textbf{iNaturalist:} This is an open-source platform for the community of Naturalists who upload observations for various species with records containing images, audio, and textual descriptions. We select observations with the flags: \texttt{Verifiable}, \texttt{Research Grade}, and \texttt{Has Sounds} to maximize data quality and completeness. This provides us with over $450$k geotagged audios. To create a relatively balanced dataset with audio from different crowd-sourced platforms, we first only retain the species with at least $100$ samples in our dataset. Then, we conduct round-robin random sampling of the observations, starting from the species with the lowest count and iteratively increasing the sample size until we reach our desired number of samples: $120$k from $611$ species. 
Finally, after a quality control filtering procedure, \textit{iNaturalist} contributes $114\,603$ audios.\\
\\
\textbf{YFCC100M:} YFCC100M is a publicly available, large multimedia dataset containing over $99$ million images and around $0.8$ million videos. This data is collected from the crowd-sourced platform \textit{Flickr}. However, among the $0.8$ million videos, only around $100$k videos are found to be geotagged. Therefore, in our dataset, we extract audio from these geotagged videos only, contributing an additional $96\,452$ audio samples. \\
\\
\textbf{Radio Aporee:} In our dataset, we also include the geotagged audios from the \textit{SoundingEarth} dataset~\citep{heidler2023self}, which was built from the crowd-sourced platform hosted by the project \textit{Radio Aporee::Maps}. This dataset contains field recordings of different types of audio from urban, rural, and natural environments. 
The \textit{SoundingEarth} dataset contributes $49\,284$ audio samples.\\
\\
\textbf{Freesound:} This is another commonly used platform for crowd-sourced audio containing field recordings from diverse acoustic environments. \textit{Freesound} contributes a total of $48\,680$ audio samples.

\begin{table}[h!]
\begin{tabular}{l|r|r|r|r|r}
\hline
split & iNaturalist & yfcc & aporee & freesound & total \\
\hline
train      & $108\,753$      & $92\,055$      & $46\,893$  & $46\,318$ & $294\,019$   \\
val & $1\,851$        & $1\,565$       & $797$    & $787$ & $5\,000$       \\
test       & $3\,999$        & $2\,832$       & $1\,594$   & $1\,575$ & $10\,000$ \\
\hline
total      & $114\,603$      & $96\,452$                           & $49\,284$                       & $48\,680$ & $309\,019$ \\
\end{tabular}
\caption{Distribution of \textit{GeoSound} Dataset Across Splits and Audio Sources.}
\label{tab:data_split_distro}
\end{table}

\subsection{Data Split Strategy}
We split our dataset to mitigate potential data leakage between data with similar locations in the training and validation/test sets. The distribution of data across training/validation/test sets and audio sources is given in Table ~\ref{tab:data_split_distro}. Our data split strategy on \textit{GeoSound} dataset is described as follows:
\begin{enumerate}
    \item We divide the world into $1$\textdegree~$\times$~$1$\textdegree~non-overlapping cells. This corresponds to the cell size of about $111$km~$\times$~$111$km. 
    \item We only select the cells with at least $25$ observations. The cells that do not pass this threshold are saved to be included in the train split of our dataset.
    \item Based on the number of observations in each cell selected in step $2$, we categorize them into three data density categories: \textit{high}, \textit{medium}, and \textit{low} based on the $0.33$ quantile and $0.66$ quantile of the overall sample count from step $2$.
    \item For each category obtained in step $3$, we randomly select $10\%$ of cells to be held out for validation and test splits.
    \item From the held-out cells obtained in step $4$, we randomly sample $40\%$ into validation split and the rest into test split.
    \item For the validation/test split, $5000/10000$ samples are randomly selected, matching the audio-source distribution of the train split.
\end{enumerate}

\section{Uncertainty Estimates}
One of the advantages of PSM is that uncertainty estimates are automatically provided with representations of samples. After a sample is encoded, the $\sigma$ associated with the distribution predicted by our framework represents its inherent uncertainty for any audio or satellite imagery. In Figure~\ref{fig:std_collage}, we present sets of samples with high uncertainty and low uncertainty for \textit{Bing} satellite imagery in our test set. 

\twocolumn[{%
\renewcommand\twocolumn[1][]{#1}%
\begin{center}
\includegraphics[height=0.3\paperheight,width=0.9\linewidth]{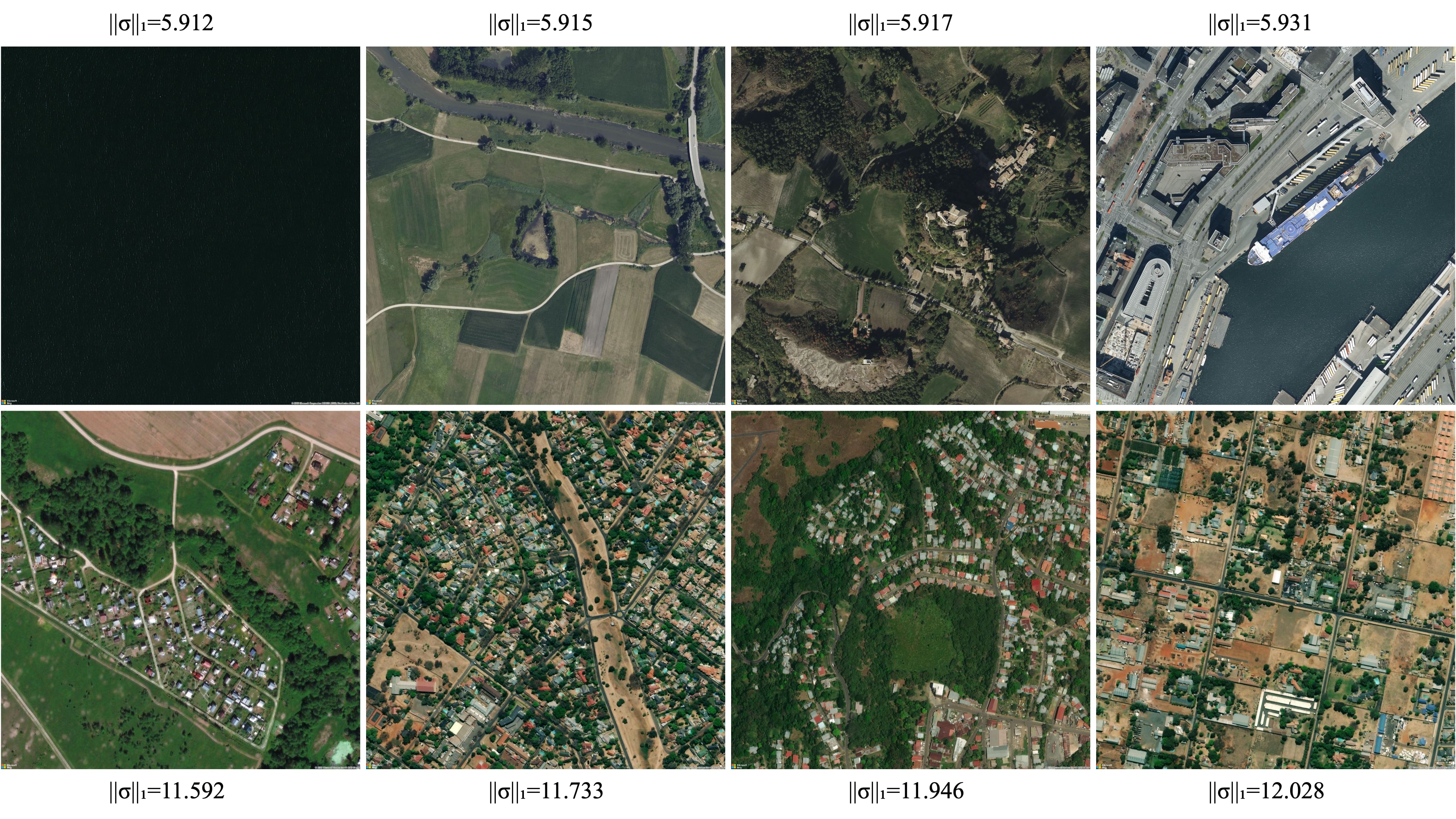}
    
    \captionof{figure}{Uncertainty estimates are reflected by the $||\sigma||_{1}$ of selected samples from our \textit{Bing} satellite imagery test set. These estimates are obtained from embeddings generated by our best-performing model trained on \textit{Bing} imagery, without any additional metadata.}
    \label{fig:std_collage}
\end{center}
\begin{center}
    \centering
    \captionsetup{type=figure}
    \includegraphics[width=\linewidth]{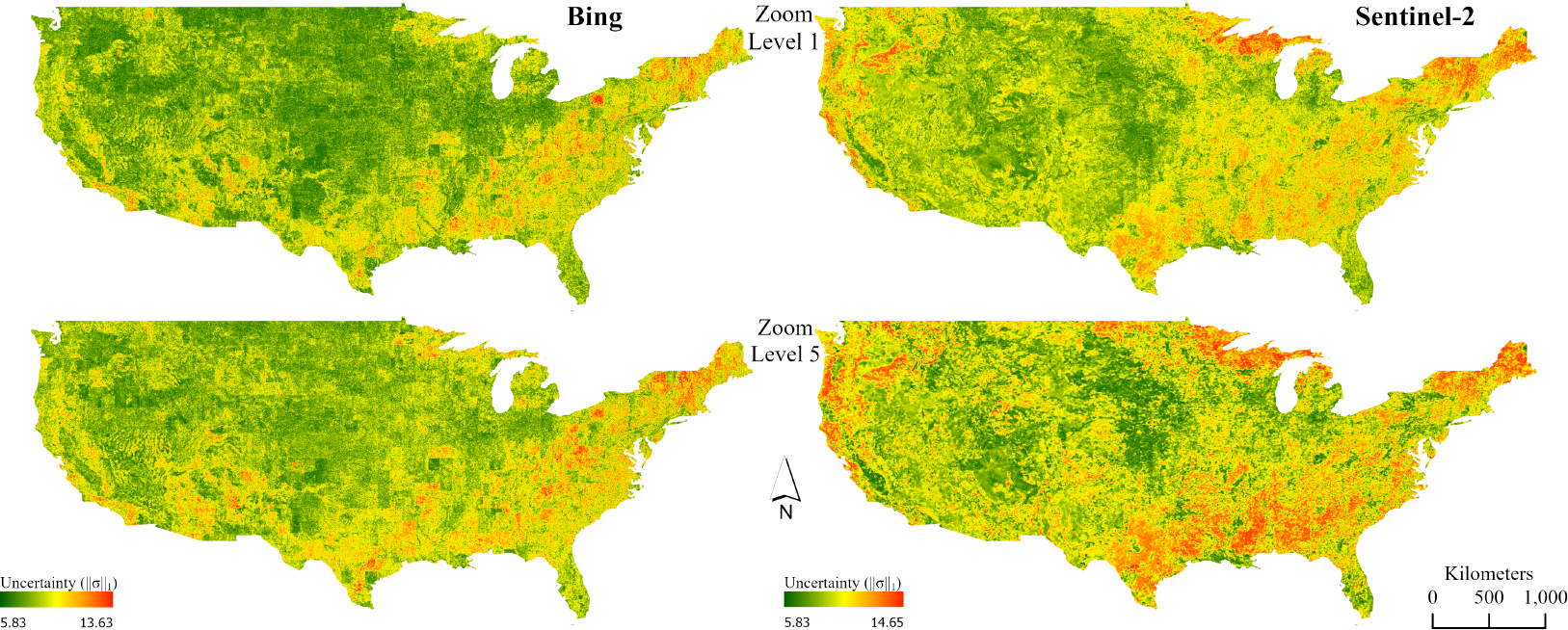}
    \captionof{figure}{Uncertainty map of the satellite image embeddings for the USA. Uncertainty at each location is approximated as the $||\sigma||_{1}$ of the probabilistic embeddings obtained from our best-performing model trained with \textit{Bing} and \textit{Sentinel-2} imagery, respectively, without any metadata.}
    \label{fig:uncertain_map}
\end{center}%
}]

The embedding dimension of our probabilistic embeddings is large ($512$); therefore, in these examples, uncertainty estimates are represented through $\lVert \sigma \rVert _{1}$ for each sample. We observe that samples with low uncertainty have fewer visible concepts captured in them, suggesting less ambiguity in the types of potential sounds that could be heard at the location. Conversely, for samples with high uncertainty, we usually find denser geographic areas where one would expect to hear multiple types of sounds, leading to higher ambiguity in soundscape mapping.

We also present country-scale uncertainty maps of the USA using PSM's satellite embeddings from both \textit{Bing} and \textit{Sentinel-2}. These maps are shown for zoom levels $1$ and $5$ in Figure~\ref{fig:uncertain_map}. From this figure, we observe that the overall distribution of uncertainty tends to be lower for zoom level $1$ compared to zoom level $5$. This result is expected because imagery at zoom level $5$ covers a larger geographic area, potentially capturing a greater diversity of soundscapes and leading to higher uncertainty in our probabilistic embedding space. Furthermore, a closer examination of the uncertainty values reveals that uncertainty estimates for \textit{Sentinel-2} image embeddings are relatively higher across more locations in the region compared to \textit{Bing} image embeddings. This is expected because for a similar image size, a \textit{Sentinel-2} image with $10$m Ground Sampling Distance (GSD) covers a larger area compared to a \textit{Bing} image with $0.6$m GSD used in our study.

\section{Metadata Dependency}
We provide a comparison of our framework's performance without using metadata during evaluation, as shown in Table \ref{tab:metadata_dependency}. As observed, the performance of PSM with text added to query embeddings, without including any other metadata, remains comparable to the best setting, which includes both text and metadata during retrieval. Using metadata without text in the query improves performance compared to not using either. This highlights the importance of encoding the dynamic nature of the soundscape based on location (latitude and longitude), season (month), and hour of the day (time). It is worth noting that, as shown in Table 4 of the main paper, which presents an ablation study of different metadata components, \textit{audio source} is the most important metadata. Therefore, during inference, the metadata type \textit{audio source} can be inferred by understanding the unique nature of sounds from each platform of audio data used to curate the dataset, as described in Section \ref{sec:data} of this supplemental material.

\begin{table}[]
\begin{tabular}{ccc|cc}
\hline
meta/train & text/eval & meta/eval & R@10\% & MdR \\
\hline
\multicolumn{5}{c}{\textbf{GeoSound-Bing}} \\
\hline
\xmark & \xmark & \xmark & 0.423 & 1401 \\
\underline{\cmark} & \underline{\xmark} & \underline{\xmark} &  \underline{0.425}   & \underline{1359}    \\
\cmark & \xmark &  \cmark &  0.828  & 261 \\
\underline{\cmark} & \underline{\cmark} & \underline{\xmark}  & \underline{0.776}   & \underline{213}   \\
\cmark & \cmark  & \cmark & 0.901 & 113 \\
\hline

\multicolumn{5}{c}{\textbf{GeoSound-Sentinel2}} \\
\hline
\xmark & \xmark & \xmark & 0.474 & 1101 \\
\underline{\cmark} & \underline{\xmark} & \underline{\xmark} & \underline{0.464}    &  \underline{1144}   \\
\cmark & \xmark &  \cmark & 0.802  &  294  \\
\underline{\cmark} & \underline{\cmark} & \underline{\xmark}  &  \underline{0.730}  & \underline{256}   \\
\cmark & \cmark  & \cmark & 0.872 & 142 \\
\hline

\multicolumn{5}{c}{\textbf{SoundingEarth}} \\
\hline
\xmark & \xmark & \xmark & 0.514  & 547 \\
\underline{\cmark} & \underline{\xmark} & \underline{\xmark} &  \underline{0.514}   &   \underline{550}  \\
\cmark & \xmark &  \cmark & 0.563  & 454   \\
\underline{\cmark} & \underline{\cmark} & \underline{\xmark}  & \underline{0.651}   & \underline{315}   \\
\cmark & \cmark  & \cmark & 0.690 & 264 \\
\hline
\end{tabular}
\caption{Evaluation of PSM under different settings for the datasets: GeoSound-Bing, GeoSound-Sentinel2, and SoundingEarth. The results are Recall@10\% (R@10\%) and Median-Rank (MdR) for Image-to-Audio retrieval with satellite image at zoom level 1. Results highlighted with \underline{underlined} text represent the evaluation of PSM without using any metadata during inference, for comparison with other evaluations presented in the main paper.}
\vspace{-20pt}
\label{tab:metadata_dependency}
\end{table}

\section{Soundscape Maps}
In Figure \ref{fig:sentinel_usa}, we present examples of country-scale soundscape maps over the USA. These maps were generated using our best-performing model trained on \textit{Sentinel-2} imagery without any metadata. In this demonstration, we utilize \textit{Sentinel-2} imagery covering the USA at zoom-level 1. In the figure, for the text query \textit{``Sound of animals on a farm''}, high activation is observed primarily in non-urban areas across the USA. Conversely, for the text query \textit{``Sound of machines in a factory''}, higher activation is concentrated in urban areas near cities, with minimal activation in forested and rangeland regions. The use of PSM trained on freely available \textit{Sentinel-2} imagery enables the creation of global-scale soundscape maps.

In Figure \ref{fig:multiscale_maps}, we showcase multi-scale soundscape mapping across various geographic regions in the USA. Our objective is to investigate how embeddings and associated similarity scores change with variations in imagery zoom level and imagery source. We generate soundscape maps using \textit{Sentinel-2} satellite image embeddings computed from imagery at zoom levels 1 and 5. To illustrate, we randomly select an audio sample from the \textit{cow} class in the ESC-50 dataset~\cite{piczak2015dataset} as an example audio query. For the text queries we select \textit{``Sound of children playing in a park''} and \textit{``Sound of machines in a factory''}. For each queries, we analyze the corresponding soundscape maps generated at different zoom levels. In Figure~\ref{fig:multiscale_maps}, each soundscape map is accompanied by a land cover map \cite{landcover} of the respective region for reference.

As observed in Figure~\ref{fig:multiscale_maps}, geographic regions expected to be related to the query demonstrate high similarity scores. For example, for the audio query described by the audio class \texttt{cow}, we can see that urban regions around cities like \textit{Memphis} and \textit{Toledo} have low similarity scores, while rural areas (with greater potential to contain farm animals) exhibit high similarity scores. Similarly, for the text query related to the sound of children playing in a park, as expected, we observe high similarity scores around cities where one would expect to find city parks.

We also observe that for the same query and geographic region, the distribution of similarity scores varies between the two zoom levels. In Figure~\ref{fig:multiscale_maps}, at zoom level 1, the generated maps appear to be more spatially fine-grained compared to maps generated using satellite imagery at zoom level 5, which appear coarser. Although the number of geolocations and their corresponding satellite imagery is the same for maps at both zoom levels, the coverage area for an image at a higher zoom level is larger. This results in a slower change of high-level visual appearance between the points in the region, leading to closer similarity scores between local points and ultimately producing soundscape maps with lower resolution. This phenomenon highlights the inherent trade-off in spatial resolution when using different zoom levels for soundscape mapping. This suggests that if we prioritize soundscape maps that retain the semantics of audio concepts at the expense of fine-grained localization capability, we can use satellite imagery at a higher zoom level, which requires fewer images to cover a region of interest. Conversely, for tasks requiring spatially fine-grained soundscape maps, satellite imagery at a lower zoom level may be preferred. This trade-off is fundamental to the multi-scale mapping capability of our framework, Probabilistic Soundscape Mapping (PSM).

\twocolumn[{%
\renewcommand\twocolumn[1][]{#1}%
\begin{center}
    \centering
    \captionsetup{type=figure}
    \includegraphics[height=0.75\paperheight,width=0.85\linewidth]{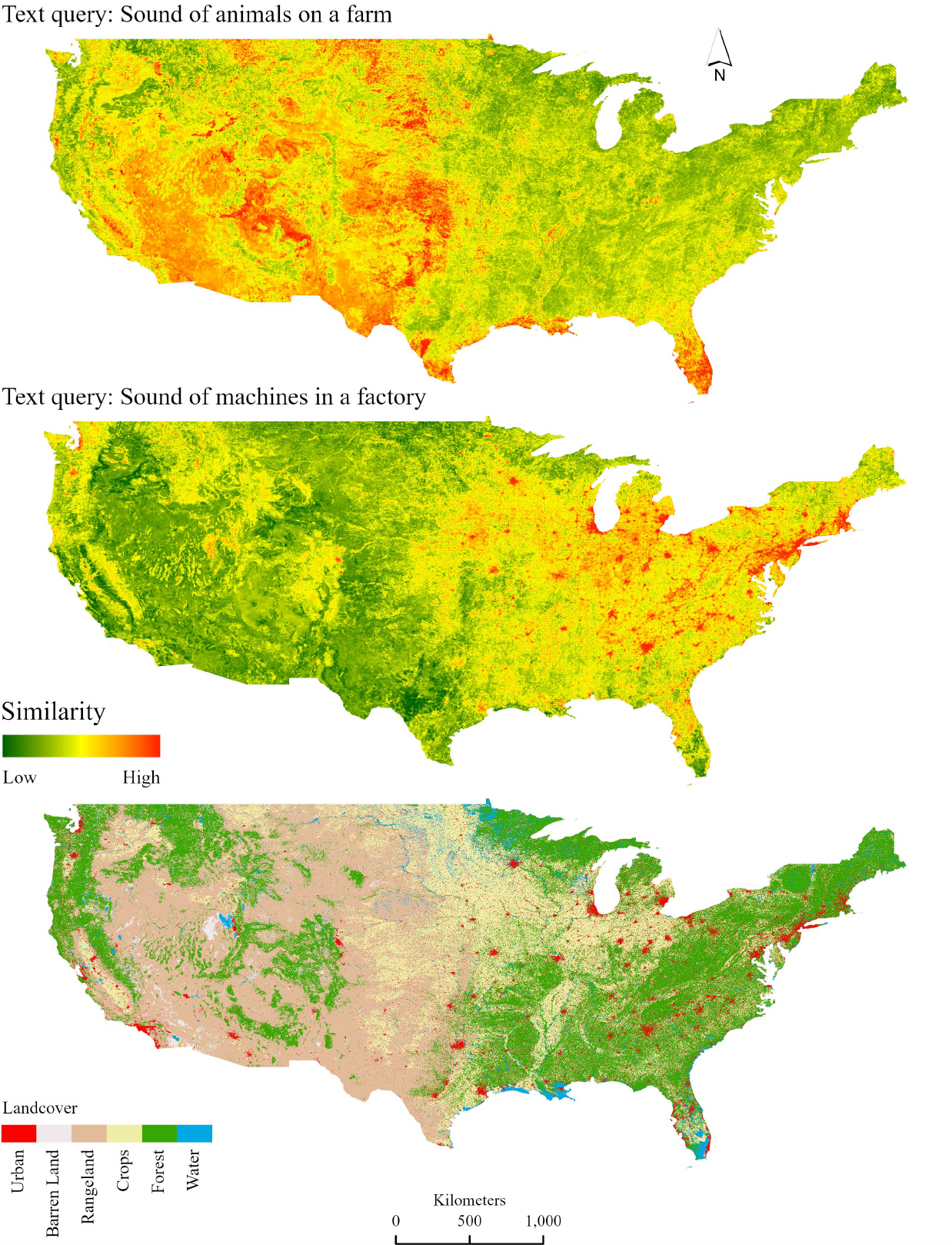}

    \caption{Two soundscape maps of the continental United States, generated from \textit{Sentinel-2} image embeddings, accompanied by a land cover map for reference \cite{landcover}.}
   
   \label{fig:sentinel_usa}
\end{center}%
}]


\twocolumn[{%
\renewcommand\twocolumn[1][]{#1}%
\begin{center}
    \centering
    \captionsetup{type=figure}
    \includegraphics[height=0.75\paperheight,width=0.85\linewidth]{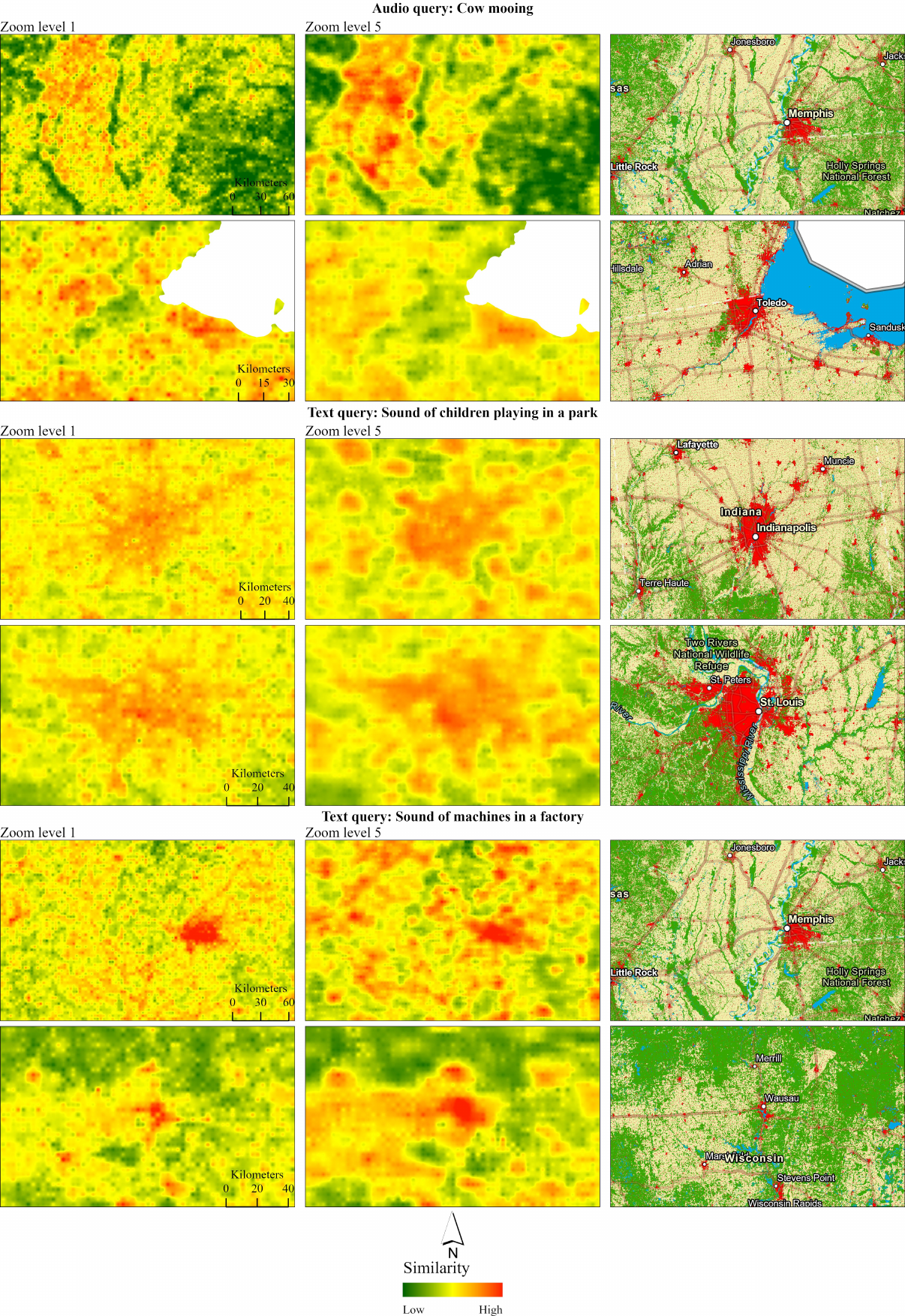}

    \captionof{figure}{Soundscape maps over smaller geographic areas, computed using similarity scores between respective queries and embeddings from \textit{Sentinel-2} satellite imagery at two zoom levels.}
   
    \label{fig:multiscale_maps}
\end{center}%
}]


\end{document}